\documentstyle[12pt]{article}

\begin{document}
\begin{titlepage}
\title{STATIC SPHERICALLY SYMMETRIC SOLUTIONS TO EINSTEIN-MAXWELL-DILATON
FIELD EQUATIONS IN D DIMENSIONS}
\author{Metin G{\" u}rses and Emre Sermutlu\\
{\small Department of Mathematics, Faculty of Science} \\
{\small Bilkent University, 06533 Ankara - Turkey}\\
{\small email:gurses@fen.bilkent.edu.tr}}
\maketitle
\begin{center}
To appear in Classical and Quantum Gravity , 1995
\end{center}
\begin{abstract}
We classify the spherically symmetric solutions of the
Einstein Maxwell Dilation
field equations in $D$-dimensions and find some exact solutions of the
string theory at all orders of the string tension parameter. We also show
the uniqueness of the black hole solutions of this theory in static axially
symmetric spacetimes.
\end{abstract}
\end{titlepage}
\section{Introduction}
Recently we observe an increase of interest in constructing
exact solutions of string theory \cite{TST}-\cite{MAN}.
There are several ways of such a construction \cite{TST}.
One of these is to start with an exact solution of the leading
order field equations (low energy limit of string theory)
and showing that this solution solves also the string equations at
all orders of the inverse power of the string tension parameter.
Plane wave spacetimes with appropriate gauge and scalar fields
have been shown to be exact solutions of the string theory
\cite{GUV}-\cite{HOR}. In these solutions the higher order
terms in the field equations dissapear due to the special form
of the plane wave metrics. Lewi-Civita-Bertotti-Robinson
spacetimes in four
dimensions is also an exact solution of the string theory
\cite{LOW},\cite{MET}.
In this solution the higher order terms do not dissapear but give some
algebraic constraints among the constants of solutions \cite{MET}.

To construct exact solutions in this
approach one must first study the low energy limit of the string theory
with some symmtery assumptions like the spherical symmetry. Among these
solutions one searches for those satisfying the field equations at all
orders. Hence we need a classification of the solutions under a symmetry
of the low energy limit of the string theory. In this work ,instead of
working directly with the low energy limit of the string theory
we consider the Einstein Maxwell Dilaton field theory with arbitrary
dilaton coupling constant $\alpha$. For some fixed values of $\alpha$ this
theory reduces to the low energy limit of the string theory and
Kaluza Klein type of theories.

With the assumption of a symmetry , static spherical symmetry
for instance , some of the solutions may describe black hole space
times.
Nonrotating black hole solutions of this theory ,
in spherically symmetric spacetimes , found so far
\cite{GIB}-\cite{DF1}  carry
mass $M$ , electric charge (or magnetic charge) $Q$ and a scalar
charge $\Sigma$. Out of these parameters only two of them are
independent .
In 4-dimensions we have the Garfinkle-Horowitz-Strominger (GHS)
solution \cite{GAR}. Gibbons and Maeda
(GM) \cite{GIB} have given black hole solutions
in arbitrary
$D$-dimensions. In four dimensions GM solution describes the region
$r>r_{1}$ of the GHS metric. In this work we find and classify the
solutions of the dilatonic Einstein-Maxwell (and hence of the low energy limit
of the
string theory) for static and spherically symmetric space-times.

Recently \cite{RAH} the most general asymptotically flat
spherically symmetric static
solution of the low energy limit of string theory in four dimensions
has been found. The uniqueness of the GHS-GM black hole solutions
in four dimensional static spherically spacetimes is manifest in this work.
Following this work
we find the most general  $D$ dimensional spherically symmetric static
solution of the Einstein Maxwell Dilaton field theory.

The uniquenes of these black hole solutions with the assumed
symmetry ($D$ dimensional spherical symmetry) is understood ,
becouse the solutions of the field equations
are completely classified. On the other hand without solving the filed
equations it is desirable to know the existance of other black hole
solutions under relaxed symmetries. Here we show that GHS-GM metric
in four dimensions is the unique asymptotically flat
static axially symmetric
black hole solution of the Einstein Maxwell Dilaton filed theory.

In this work we assume $D$ dimensional static spherically symmetric
spacetimes. We first write , in section 2 ,the geometrical quantitites
like riemann
and ricci tensors in terms of the volume form of the sphere $S^{D-2}$
or in terms of its dual form. With such a representation it is relatively
easier to work with the higher order curvature terms. In the third section
we classify the static spherically symmetric solutions of the Einstein
Maxwell Dilaton field equations. In the fourth section we prove that
the GHS+GM metrics are the unique static axially symmetric
black holes of the Einstein Maxwell Dilaton field theory. In the last section
we give some theorems for the $D$ dimesional static spherically symmetric
spacetimes which  are crucial to find the exact solutions of the string
theory . In this section we give some possible exact solutions of string
theory in $D$-dimensions.


\section{Static Spherically Symmetric Riemannian Geometry}

The line element of a static and spherically symmetric spacetime
is given by:

\begin{equation}
ds^{2}=-A^{2}dt^{2}+B^2dr^{2}+C^2d\Omega_{D-2}^{2}  \label{met}
\end{equation}

\noindent
where A,B,C depend only on r. $d\Omega_{D-2}^{2}$ is the
line element on $S_{D-2}$. The metric can be rewritten as
$g_{ij}=-A^{2}t_{i}t_{j}+B^{2}k_{i}k_{j}+C^{2}h_{ij}$,
where $t_{i}=\delta^{t}_{i}$, $k_{i}=\delta^{r}_{i}$,
$h_{ij}=$metric on D-2 sphere for  $ i,j\ge2,\;h_{0i}=h_{1i}=0$

\noindent
Components of the christoffel symbol are given by

\begin{eqnarray}
\Gamma^{i}_{jm}&=&-AA^{\prime}\,t^{i}(t_{j}k_{m}+t_{m}k_{j})+
AA^{\prime}\,k^{i}t_{j}t_{m}+BB^{\prime}\,k^{i}k_{j}k_{m}-
CC^{\prime}\,k^{i}h_{jm} \nonumber  \\
 & &+CC^{\prime}\,(h^{i}_{j}k_{m}+h^{i}_{m}k_{j})+\Gamma^{i}_{(s)jm}
\end{eqnarray}

\noindent
where $\Gamma_{(s)}$ is the christoffel symbol on $D-2$ sphere

\noindent
The Riemann tensor is given by

\begin{equation}
R^{i}_{jml}=\Gamma^{i}_{jl,m}-\Gamma^{i}_{jm,l}+\Gamma^{i}_{nm}\Gamma^{n}_{jl}-
\Gamma^{i}_{nl}\Gamma^{n}_{jm}
\end{equation}

\noindent
We find that

\begin{eqnarray}
R_{ijml}&=&\left(AA^{\prime\prime}-
{AA^{\prime}B^{\prime}\over  B}\right)\,t_{[i}k_{j]}t_{[m}k_{l]}
+\left(CC^{\prime\prime}-{CC^{\prime}B^{\prime}\over
B}\right)\,k_{[i}h_{j][m}k_{l]} \nonumber  \\
 & &+{AA^{\prime}CC^{\prime}\over  B^2}\,t_{[m}h_{l][j}t_{i]}-
 {C^2C^{\prime^{2}}\over  B^2} h_{i[m}h_{l]j}+C^2 R_{(s) ijml} \label{ijm}
\end{eqnarray}

\noindent
where
$R_{(s) ijml}=h_{i[m}h_{l]j}$ are the components
of the Riemann tensor on $S_{D-2}$.
Riemann tensor in (\ref{ijm}) may be rewritten as

\begin{equation}
R_{ijml}=g_{jl}S_{im}-g_{jm}S_{il}+g_{im}S_{jl}-g_{il}S_{mj}
+\eta_{2}H_{ijk\cdots n}H_{ml}^{~~k\cdots n} \label{brn}
\end{equation}

\noindent
where

\begin{equation}
S_{ij}=\eta_{0}M_{ij}+\eta_{1}k_{i}k_{j}+{1 \over  2} \eta_{3}g_{ij}
\end{equation}

\begin{equation}
M_{ij}=H_{im\cdots n}H_{j}^{m\cdots n}-{1\over {2 (D-2)}}H^{2}g_{ij}
\end{equation}

\noindent
which turns out to be
\begin{equation}
M_{ij}={(D-3)!\over {C^{2 (D-3)}}} (h_{ij}-{1\over {2 C^{2}}}g_{ij})
\end{equation}
\noindent
Here the tensor $ H_{ij...k}$ is the volume form of $S_{D-2}$, i.e.

\begin{equation}
H_{ij\cdots k}=-\sqrt{h}\, \epsilon_{ij\cdots k}
\end{equation}
\noindent
where $h$ denotes det$h_{ij}$

\noindent
Here the scalars $\eta_{i}$ are given by

\begin{equation}
\eta_{0}={{C^{2 (D-2)}}\over {(D-3)!}}\left({{A^{\prime\prime}}\over {A
B^{2}}}-
{{A^{\prime}B^{\prime}}\over {AB^{3}}}-{{C^{\prime\prime}}\over {CB^{2}}}+
{{B^{\prime}C^{\prime}}\over {B^{3}C}}\right)
\end{equation}

\begin{equation}
\eta_{1}={{A^{\prime}C^{\prime}}\over {AC}}-{{C^{\prime\prime}}\over
{C}}+{{B^{\prime}C^{\prime}}\over {BC}}
\end{equation}

\begin{equation}
\eta_{2}={{C^{2 (D-3)}}\over {(D-4)!}} \left(1-{{C^{\prime^{2}}}\over {B^{2}}}+
{{A^{\prime}C^{\prime}C}\over {A B^{2}}}-{{A^{\prime\prime}C^{2}}\over
{AB^{2}}}+
{{A^{\prime}B^{\prime}C^{2}}\over {AB^{3}}}+{{CC^{\prime\prime}}\over {B^{2}}}
-{{B^{\prime}C^{\prime}C}\over {B^{3}}}\right)
\end{equation}

\begin{equation}
\eta_{3}=-{{A^{\prime}C^{\prime}}\over {AB^{2}C}}
\end{equation}

\noindent
The Ricci tensor, Ricci scalar and Einstein tensor can be computed
from Riemann as follows:

\begin{eqnarray}
R_{ij}&=&\left[{{(D-3)!}\over {2C^{2(D-2)}}}(\eta_{0}(D-4)+
\eta_{2})+{{1}\over {B^{2}}}\eta_{1}+\eta_{3}(D-1)\right]g_{ij} \nonumber  \\
 & &+\eta_{1}(D-2)k_{i}k_{j}+[\eta_{0}(D-2)+\eta_{2}]M_{ij} \\
 & & \nonumber  \\
R&=&{{(D-3)!}\over {C^{2(D-2)}}}[\eta_{0}(D-4)(D-1)+\eta_{2}(D-2)]+
{{2 \eta_{1}(D-1)}\over {B^{2}}} \nonumber  \\ & &+\eta_{3}D(D-1)  \\
 & &  \nonumber  \\
G_{ij}&=&-\left[{{(D-3)!}\over {2C^{2(D-2)}}}(\eta_{0}(D-4)(D-2)+
\eta_{2}(D-3))+{{1}\over {B^{2}}}\eta_{1}(D-2)\right.  \nonumber  \\
 & &\left.+{{1}\over {2}}\eta_{3}(D-1)(D-2)\right]g_{ij}+
 \eta_{1}(D-2)k_{i}k_{j}\nonumber \\
 & &+[\eta_{0}(D-2)+\eta_{2}]M_{ij} \label{sch}
\end{eqnarray}

The covariant derivatives of $H_{ij...k}$ and $k_{i}$ are given as

\begin{equation}
\nabla_{l}H_{ij...m}=-\rho[(D-2)H_{ij...m}k_{l}+
k_{i}H_{lj...m}+k_{j}H_{il...m}+...+k_{m}H_{ij...l}] \label{xy1}
\end{equation}

\begin{equation}
\nabla_{i}k_{j}=\rho_{1}g_{ij}+\rho_{2}M_{ij}+\rho_{3}k_{i}k_{j}
\label{xy2}
\end{equation}

\noindent
where

\begin{eqnarray}
\rho&=&{{C^{\prime}}\over {C}} \mbox{ \hspace{4cm}  } \rho_{1}=
{{A^{\prime}C+AC^{\prime}}\over {2AB^{2}C}}  \nonumber  \\
 & &  \nonumber  \\
\rho_{2}&=&{{C^{2(D-2)}}\over {(D-3)!}} {{A^{\prime}C-AC^{\prime}}\over
{AB^{2}C}}
\mbox{\hspace{1.1cm}} \rho_{3}=-{{(AB)^{\prime}}\over {AB}}
\end{eqnarray}

In (\ref{brn}) we have given the Riemann
tensor of a static, spherically symmetric spacetime
in terms of the volume form of $S_{D-2}$ (D-2)-brane field.
We may also write the Riemann tensor in the form

\begin{equation}
R_{ijml}=g_{jl}S_{im}-g_{jm}S_{il}+g_{im}S_{jl}-g_{il}S_{mj}
+e_{2}F_{ij}F_{ml}
\end{equation}

\noindent
where
\begin{equation}
F_{ij}={{AB}\over {C^{2}}} (t_{i}k_{j}-t_{j}k_{i}) \label{d12}
\end{equation}
\noindent
Other tensors are defined similar to the previous case, i.e.

\begin{equation}
S_{ij}=e_{0}\,\widetilde{M_{ij}}+e_{1}\,k_{i}k_{j}+{1 \over  2} e_{3}\,g_{ij}
\end{equation}

\begin{equation}
\widetilde{M_{ij}}=F_{mj}F_{i}^{m}-{1\over {4}}F^{2}\,g_{ij}
\end{equation}

\begin{equation}
\widetilde{M_{ij}}={1\over {C^{2}}} (h_{ij}-{1\over {2\, C^{2}}}\,g_{ij})
\end{equation}

\noindent
The scalars are given as

\begin{eqnarray}
e_{0}&=&C^{2}-{{C^{\prime^{2}}C^{2}}\over {B^{2}}}+
{{C^{3}C^{\prime}A^{\prime}}\over {A\,B^{2}}}  \nonumber \\
e_{1}&=&\eta_{1}    \\
e_{2}&=&-C^{2} \left(1-{{C^{\prime^{2}}}\over {B^{2}}}
+{{A^{\prime}C^{\prime}C}\over {A B^{2}}}-{{A^{\prime\prime}C^{2}}\over
{AB^{2}}}+
{{A^{\prime}B^{\prime}C^{2}}\over {AB^{3}}}+{{CC^{\prime\prime}}\over {B^{2}}}-
{{B^{\prime}C^{\prime}C}\over {B^{3}}}\right) \nonumber  \\
e_{3}&=&\eta_{3}  \nonumber
\end{eqnarray}

\noindent
Notice that when $D=4$, $\widetilde{M_{ij}}=M_{ij}$, and $e_{2}=-\eta_{2}$.

The covariant derivatives of $F_{ij}$, $k_{i}$ and $t_{i}$ are given as

\begin{equation}
\nabla_{l}F_{ij}=-{{3C^{\prime}}\over {C}}F_{ij}k_{l}-
{{AC^{\prime}}\over {BC^{3}}}(t_{j}g_{il}-t_{i}g_{jl})
\end{equation}

\begin{equation}
\nabla_{i}k_{j}=\rho_{1}g_{ij}+\tilde{\rho_{2}}\widetilde{M_{ij}}+
\rho_{3}k_{i}k_{j}
\end{equation}

\begin{equation}
\nabla_{i}t_{j}={A^{\prime}\over  A} (t_{i}k_{j}+t_{j}k_{i})
\end{equation}

\noindent
where $\tilde \rho_{2}={{(D-3)!}\over {C^{2(D-4)}}}\,\rho_{2} $

\section{Solutions of the Einstein Maxwell Dilaton Field Equations}

The field equations of the Einstein Maxwell Dilaton
theory can be obtained from the following lagrangian

\begin{equation}
L={\sqrt{ -g}}\left[{R\over  2\kappa^{2}}-{4\over  (D-2)\kappa^{2}}
(\nabla \phi)^{2}-{1\over  4}\ e^{-\alpha_{e} \phi}F^{2}\right]
\end{equation}

\noindent
The field equations are

\begin{eqnarray}
G_{ij}&=&{8\over  (D-2)}\left[\partial_{i} \phi\partial_{j}\phi
-{1\over  2}(\nabla\phi)^{2} g_{ij} \right] -\kappa^{2} e^{-\alpha_{e} \phi}
\left[F_{i}^{m} F_{jm}-{1\over  4} F^{2} g_{ij} \right] \nonumber  \\
 \label{EJ}
\end{eqnarray}

\begin{equation}
\nabla_{i} (e^{-\alpha_{e} \phi } F^{ij}) = 0 , \\   \label{EI}
\end{equation}

\begin{equation}
\partial_{i} ({\sqrt -g}\, g^{ij}\, \partial_{j} \phi)
 + {(D-2) \kappa^{2} \alpha_{e} {\sqrt -g} \over  32} e^{-\alpha_{e} \phi}
 F^2 = 0  ,
 \label{EE}
\end{equation}

\noindent
where $F_{ij}$ is the Maxwell and  $\phi$ is the dilaton field.
Here $i,j=1,2,... ,D \ge 4$.

\noindent
In static spherically symmetric spacetimes, gravitational field
equations first lead to

\begin{eqnarray}
{\eta_{0}{(D-4)(D-2)!}\over {2C^{2(D-2)}}}+
{\eta_{2}{(D-3)(D-3)!}\over {2C^{2(D-2)}}}+
{\eta_{1}{(D-2)}\over {B^{2}}}+
{\eta_{3}{(D-1)(D-2)}\over {2}}& & \nonumber  \label{111}  \\
-{{4\phi^{\prime^{2}}}\over {(D-2)B^{2}}}=0 \mbox{\hspace{10cm}} & &
\end{eqnarray}

\begin{equation}
\eta_{1}(D-2)-{{8\phi^{\prime^{2}}}\over {(D-2)}}=0 \label{222}
\end{equation}

\begin{equation}
\eta_{0}(D-2)!+\eta_{2}(D-3)!-{\kappa^{2}Q^{2}e^{\alpha_{e}\phi}\over
A^{2}_{D-2}}=0
\end{equation}

\noindent
Dilaton equation is

\begin{equation}
{{8}\over {D-2}}\left[{{A}\over {B}}\,C^{D-2}\,\phi^{\prime}\right]^{\prime}-
{{\alpha_{e}\,A\,B\,\kappa^{2}\,Q^{2}\,e^{\alpha_{e}\phi}}\over {2\,
C^{D-2}\,A^{2}_{D-2}}}=0 \label{dlt}
\end{equation}

\noindent
{}From (\ref{111}) and (\ref{222}) we obtain

\begin{equation}
\left [\left ( A\,C^{d} \right)^{\prime}\,C \over  B \right ]^{\prime}=
d^{2}\,A\,B\,C^{d-1} \label{asl}
\end{equation}

\noindent
where $d=D-3$. Using the freedom in choosing the $r$ coordinate we can
let

\begin{equation}
A\,B\,C^{d-1}=r^{d-1}, \label{DE1}
\end{equation}

\noindent
Using (\ref{asl}) and (\ref{DE1}) we obtain

\begin{equation}
A^{2}\,C^{2d}=r^{2d}-2b_{1}\,r^{d}+b_{2} \label{qua}
\end{equation}

\noindent
where $b_{1}$ and $b_{2}$ are integration constants.

 A combination of the dilaton (\ref{EI}) and gravitational field
 equations (\ref{EJ}) give

\begin{equation}
d\, T^{2}- {d-1 \over  r}\,T + T^{\prime}+ {8 \over  (d+1)^{2}} \,
\phi^{\prime\, 2}=0  , \label{EQ4}
\end{equation}

\noindent
where $T$ is defined as

\begin{equation}
T={(r^{d}+c_{1})\,r^{d-1} \over  (r^{2d}-2b_{1}r^{d}+b_{2})}-
{16 \phi^{\prime} \over  (d+1)^{2} \alpha_{e}}.
\end{equation}

\noindent
Here $c_{1}$ is an integration constant
Defining now a new function $\psi(\rho)$ as follows

\begin{equation}
\alpha_{e}\,\phi^{\prime}={k_{1}\,d\,r^{d-1} \over  (r^{2d}-
2b_{1}r^{d}+b_{2})}\,\psi(\rho)
\end{equation}

\noindent
the equation (\ref{EQ4}) becomes

\begin{equation}
{r^{2d}-2b_{1}r^{d}+b_{2}\over  d\,r^{d-1}}\;{d\psi \over  d\rho}\;
{d\rho\over  dr}=(\psi+\mu)^{2}+\nu^{2} \label{dn1}
\end{equation}

\noindent
The constants are given by

\begin{eqnarray}
a&=&{(d+1)^{2}\, \alpha_{e}^{2} \over  32\,d}  \nonumber  \\
\mu&=&-(c_{1}+b_{1}) \label{mna} \\
\nu^{2}&=&a\mu^{2}-\Delta(a+1)  \nonumber
\end{eqnarray}

\noindent
Now, if we solve the auxiliary equation

\begin{equation}
{r^{2d}-2b_{1}r^{d}+b_{2}\over  d\,r^{d-1}}{d\rho\over  dr}=\rho
\end{equation}

\noindent
and insert $\rho$ above in Eq.(\ref{dn1}) , we obtain

\begin{equation}
\rho {d\psi \over  d\rho}=(\psi+\mu)^{2}+\nu^{2} , \label{EQ5}
\end{equation}

\noindent
Note that $\rho$ depends on the sign of $\Delta=b_{1}^{2}-
b_{2}$. Hence $\phi$ can be found from $\psi$ by

\begin{equation}
\phi={k_{1}\over  \alpha_{e}}\,\int{\psi(\rho)\over  \rho}d\rho+\phi_{0}
\end{equation}

\noindent
Where $k_{1}={2\,a\over  a+1}$ and $\phi_{0}$ is an arbitrary constant.
The metric function $C$ is connected
to $\phi$ as

\begin{equation}
{C^{\prime} \over  C}={(r^{d}+c_{1})r^{d-1} \over  (r^{2d}-
2b_{1}r^{d}+b_{2})}-{16 \phi^{\prime} \over  (d+1)^{2} \alpha_{e}}
\end{equation}

\noindent
This gives us

\begin{equation}
\ln ({C\over  c_{0}})^{d}=\left(\int{u+b_{1}+c_{1}\over
u^{2}-\Delta}du\right)-
{\alpha_{e}\over  2a}\phi
\end{equation}

\noindent
Where $u=r^{d}-b_{1}$ and $c_{0}$ is an arbitrary integration constant.

\noindent
Metric functions $A$ and $B$ can be found from $C$ through the
equations (\ref{DE1}) and (\ref{qua})

\noindent
We have three different cases according to the sign of $\Delta$.

\begin{eqnarray}
\mbox{Case 1   }&\Delta>0&\ln\rho(r)={1\over
r_{1}^{d}-r_{2}^{d}}\ln\left({r^{d}-
r_{1}^{d}\over  r^{d}-r_{2}^{d}}\right) \nonumber  \\
\mbox{Case 2   }&\Delta=0&\ln\rho(r)=-{1\over  r^{d}-r_{3}^{d}} \\
\mbox{Case 3   }&\Delta<0&\ln\rho(r)={1\over
\sqrt{-\Delta}}\left[\arctan\left({r^{d}-
b_{1}\over  \sqrt{-\Delta}}\right)-{\pi\over  2}\right] \nonumber
\end{eqnarray}

\noindent
Where
$r_{1}^{d}=b_{1}+\sqrt{\Delta},\;r_{2}^{d}=b_{1}-\sqrt{\Delta},\;r_{3}^{d}=
b_{1}$.

\noindent
The integration constants $c_{0}$ and $\phi_{0}$ are determined by taking
the asymptotic behaviour of the functions $C(r)$ and $\phi(r)$ to be

\begin{equation}
\lim_{r \rightarrow \infty} \phi =\phi_{0}=0
\end{equation}

\begin{equation}
\lim_{r \rightarrow \infty} {C(r)\over  r}=c_{0}=1
\end{equation}

\noindent
To determine the remaining integration  constants, we restrict the asymptotic
behaviour of the metric, scalar field and the tensor field $F_{ij}$ as follows:

\begin{eqnarray}
\lim_{r \rightarrow \infty}\,r^{d}\,( A^{2}-1)&=&- {2\kappa^{2} \,M\,
\over  A_{d+1}\, (d+1)}  \nonumber    \\
\lim_{r \rightarrow \infty} r^{d+1}\phi^{\prime}&=
&-{\kappa \sqrt{d+1} \over  2A_{d+1}}\, \Sigma  \label{fiz}   \\
\lim_{r \rightarrow \infty} r^{d+1}\,F_{tr}&=&{Q \over  A_{d+1}} \nonumber
\end{eqnarray}

\noindent
where $M$ , $\Sigma$ , and $Q$ are the mass , dilaton and electric charges
respectively \cite{GIB}. From the definition of these constants it is clear
that we are interested in the asymptotically flat solutions of the Einstein
Maxwell Dilaton field equations

{\bf CASE 1}

\begin{equation}
\Delta>0
\end{equation}

\noindent
Which means there are two roots to the equation (\ref{qua}).
According to the sign of $\nu^{2}$ we have three distinct solutions

{\bf Type 1}

\begin{eqnarray}
\nu^{2}&<&0 \nonumber  \mbox{\hspace{2cm}} \lambda^{2}=-\nu^{2}\\
\psi&=&{\lambda-\mu+c_{2}(\lambda+\mu)\rho^{2\lambda} \over
1-c_{2}\rho^{2\lambda}} \nonumber
\end{eqnarray}

\noindent
The metric functions become

\begin{eqnarray}
A^{2}&=&{(r^{d}-r_{1}^{d})(r^{d}-r_{2}^{d}) \over  C^{2d}}\mbox{\hspace{1cm}}
B^{2}={r^{2d-2}\,C^{2} \over  (r^{d}-r_{1}^{d})(r^{d}-r_{2}^{d})} \\
C^{d}&=&(r^{d}-r_{2}^{d}) \left(1-c_{2}\,\rho^{2\lambda}\over  1-
c_{2}\right)^{k_{2}}\rho^{k_{3}}  \\
\end{eqnarray}

\noindent
Dilaton field is given as

\begin{equation}
e^{\alpha_{e}\, \phi}=\left [{ (1-c_{2})\,\rho^{\lambda-\mu} \over  1-
c_{2}\,\rho^{2\lambda} } \right ]^{k_{1}}
\end{equation}

\noindent
The constants are given by

\begin{equation}
k_{2}={1\over  (a+1)},\mbox{\hspace{2cm}}  k_{3}={(r_{1}^{d}-r_{2}^{d})\over
2}-
{(\nu+a\mu)\over  (a+1)}
\end{equation}

\noindent
Undetermined integration constants are $r_{1},\,r_{2},\,c_{1}$ and
$c_{2}$. From the boundary condition (\ref{fiz}) we find that

\begin{eqnarray}
2\, M&=&\left[{1+c_{2}\over  1-c_{2}}\, \lambda+ a\,\mu \right] e_{1} \nonumber
 \\
\Sigma&=&\left[{1+c_{2}\over  1-c_{2}}\, \lambda-\, \mu \right] e_{2}
\label{ala}   \\
Q&=&\lambda \,{\sqrt{c_{2}}\over  1-c_{2}} e_{3} \nonumber
\end{eqnarray}

\noindent
The constants $e_{i}$ are given by

\begin{eqnarray}
e_{1}&=&{2\, A_{d+1}\,(d+1)\over  (a+1)\kappa^{2}} \nonumber  \\
e_{2}&=&{A_{d+1}\,\alpha_{e}\,(d+1)^{{3\over  2}} \over  8\,(a+1)
\kappa} \label{EQ9}
  \\
e_{3}&=&{2\,A_{d+1}\over  \kappa}\sqrt{(d+1)\,d\over  (a+1)} \nonumber
\end{eqnarray}

\noindent
Note that, we have four integration constants ($c_{1},\,c_{2},\,r_{1}^{d}$
and $r_{2}^{d}$) but there exists only three equations to determine them.
Also note that $c_{1}$ does not appear in the solution directly, so we
have a freedom in $c_{1}$. (
$c_{2}\neq1$) In order to complete the solution,we need to determine the
integration constants in terms of the physical parameters $M$,
$\Sigma$ and $Q$. Let us define some auxiliary variables to solve the
set of algebraic equations (\ref{ala})

\begin{eqnarray}
T_{1}&=&{1\over  a+1} ({a\, \Sigma \over  e_{2}}+{2\, M \over  e_{1}})
\mbox{\hspace{1cm}} T_{2}={1\over  a+1}\, ({2\, M \over  e_{1}}-
{ \Sigma \over  e_{2}}) \nonumber  \\
T_{3}&=&{\sqrt {1-{4\,Q^{2} \over  e_{3}^{2}\,T_{1}^{2}}}}
\end{eqnarray}

\noindent
Then the integration constants are

\begin{eqnarray}
\lambda&=& T_{3}T_{1} \nonumber \\
\mu&=&T_{2}  \label{cri} \\
c_{2}&=&{1-T{_3}\over  1+ T_{3}} \nonumber
\end{eqnarray}

\noindent
Also

\begin{equation}
\Delta={a\mu^{2}+\lambda^{2}\over  a+1}
\end{equation}

\noindent
The reality of $T_{3}$ imposes

\begin{equation}
M+g\,\Sigma \ge s\, \vert Q \vert \label{inq}
\end{equation}

\noindent
where $g$ and $s$ are given by

\begin{equation}
g={\alpha_{e}\, (d+1)^{3/2} \over  4d\, \kappa} \;\; , \;\; s=
{1\over  \kappa}{\sqrt {(d+1)(a+1)\over  d}} \label{ggs}
\end{equation}

\noindent
Such an inequality has been found by Gibbons and Wells for D=4.

\noindent
When $\Delta>0$ we have two roots to the equation (\ref{qua}). In general these
roots are the singular
points of the space-time. If the integration constants satisfy some
additional constraints one of these roots becomes regular. An invariant
of the space-time is the scalar curvature is given by

\begin{equation}
R=A_{1}{\rho^{z_{1}}\left[-\mu+\lambda+{2 c_{2} \lambda \rho^{2\lambda}\over
1-
c_{2}\rho^{2\lambda}} \right]^{2}\over  [(r^{d}-r_{1}^{d})(r^{d}-
r_{2}^{d})]^{1+{1\over  d}}(1-c_{2}\rho^{2\lambda})^{z_{2}}}+
A_{2}{\rho^{z_{1}+2\lambda}\left({1-c_{2}\over  1-c_{2}\rho^{2\lambda}}\right)^
{z_{2}+2}\over [(r^{d}-r_{1}^{d})(r^{d}-r_{2}^{d})]^{1+{1\over d}}} \label{kor}
\end{equation}

\noindent
where

\begin{eqnarray}
z_{1}&=&{2\over d(a+1)}(a \mu+\lambda) \nonumber \\
z_{2}&=&{2\over d(a+1)} \nonumber \\
A_{1}&=&{8\over d+1}{k_{1}^{2}d^{2}\over \alpha_{e}^{2}} \\
A_{2}&=&{\kappa^{2}\over 2}{(D-4)\over (D-2)} \nonumber
\end{eqnarray}

\noindent
As  $r\rightarrow r_{1}$ we have a singularity unless
we choose $\mu=\lambda$. This choice, by the utility of
eqn (\ref{mna}), gives $\mu=\lambda={r_{1}^{d}-r_{2}^{d}\over 2}$.
Inserting this in equation (\ref{kor}), we find that as $r\rightarrow r
_{1}$,

\begin{equation}
R=\tilde{A_{1}}\,(r^{d}-r_{1}^{d})+\tilde{A_{2}}
\end{equation}

\noindent
where $\tilde{A_{1}}$ and $\tilde{A_{2}}$ are constants ,
so the horizon is regular. The choice $\mu=\lambda$ gives $T_{2}=T_{1}T_{3}$ by
equaton (\ref{cri}), which means

\begin{equation}
Q^{2}={8 d\Sigma \over  \alpha_{e}} \left [ {2 \kappa \over (d+1)^{3\over 2}}
M+{8 (a-1)\over \alpha_{e} (d+1)^{2}} \Sigma \right ]
\end{equation}

\noindent
The connection between physical parameters and integration constants is

\begin{eqnarray}
2M&=&{r_{1}^{d}-r_{2}^{d}\over 2}\left({1+c_{2}\over 1-c_{2}}+a\right) e_{1}
\nonumber \\
\Sigma&=&{r_{1}^{d}-r_{2}^{d}\over 2} {2c_{2}\over 1-c_{2}} e_{2}   \\
Q&=&{r_{1}^{d}-r_{2}^{d}\over 2}{\sqrt{c_{2}}\over 1-c_{2}} e_{3}  \nonumber
\\
\end{eqnarray}

\noindent
The metric is

\begin{equation}
ds^{2}=-{(r^{d}-r_{1}^{d})(r^{d}-
r_{2}^{d})\over C^{2d}}dt^{2}+{r^{2d-2}C^{2}\over (r^{d}-
r_{1}^{d})(r^{d}-r_{2}^{d})}dr^{2}+C^{2}d\Omega_{D-2}^{2} \label{l11}
\end{equation}

\noindent
where

\begin{equation}
C^{d}=(r^{d}-r_{2}^{d})\left(1+{\Sigma/e_{2}\over r^{d}-
r_{2}^{d}}\right)^{k_{2}} \label{ME3}
\end{equation}

\begin{equation}
e^{\alpha_{e}\, \phi}=\left(1+{\Sigma/e_{2}\over r^{d}-
r_{2}^{d}}\right)^{-k_{1}}
\end{equation}

\begin{equation}
F_{tr}={Q\over A_{d+1} r^{d+1}}
\end{equation}
and ${r_{1}^{d}-r_{2}^{d}\over 2}=T_{2}$.

If we choose the integration constant $r_{2}$ as zero, these black hole metrics
are identical
with the metrics given by Gibbons and Maeda \cite{GIB}

\noindent
Although the extreme limit ($\Delta=0$) in genereal
is going to be studied in Case II we obtain the extreme case of the
above solution by setting $c_{2}=(r_{2}/r_{1})^{d}$ and $r_{2}=r_{1}$.
Then we obtain

\begin{eqnarray}
2M&=&r_{1}^{d} e_{1} \nonumber \\
\Sigma&=&r_{1}^{d} e_{2}   \\
2Q&=&r_{1}^{d} e_{3}  \nonumber  \\
\end{eqnarray}

{\bf Type 2}

\begin{eqnarray}
\nu^{2}&>&0 \nonumber   \\
\psi&=&\nu\,\tan(c_{2}+\nu \ln \rho)-\mu
\end{eqnarray}

\begin{equation}
\int {\psi(\rho)\over \rho}d\rho=-\ln\left[\cos(c_{2}+\nu\ln\rho)\right]-
\mu \ln\rho+c_{3}
\end{equation}

\noindent
After similar steps as the previous type, we arrive at the solution

\begin{eqnarray}
A^{2}&=&{(r^{d}-r_{1}^{d})(r^{d}-r_{2}^{d}) \over C^{2d}}\mbox{\hspace{1cm}}
B^{2}={r^{2d-2}\,C^{2} \over (r^{d}-r_{1}^{d})(r^{d}-r_{2}^{d})} \\
C^{d}&=&(r^{d}-r_{2}^{d})\,\rho^{{r_{1}^{d}-r_{2}^{d}\over 2}-
{\mu\,k_{1}\over 2}}\left[\cos(c_{2}+
\nu\ln\rho)\over \cos{c_{2}}\right]^{k_{2}}
\end{eqnarray}

\noindent
Scalar field is given as

\begin{equation}
e^{\alpha_{e}\, \phi}=\left [{\cos{c_{2}} \over \rho^{\mu}\,\cos({c_{2}+
\nu\ln\rho)}} \right ]^{k_{1}}
\end{equation}

\noindent
Physical parameters are found using (\ref{fiz})

\begin{eqnarray}
2M&=&(a\mu+\nu\tan c_{2})\, e_{1} \nonumber  \\
\Sigma&=&(-\mu+\nu\tan c_{2}) \,e_{2}   \\
Q&=&{\nu\over \cos c_{2}} \,e_{3} \nonumber
\end{eqnarray}

\begin{eqnarray}
\mu&=&T_{2} \nonumber  \\
\sin c_{2}&=&{e_{3}\,T_{1}\over Q} \\
\nu \tan c_{2}&=&T_{1}  \nonumber
\end{eqnarray}

\noindent
The condition $|\sin c_{2}|<1$ imposes

\begin{equation}
M+g\,\Sigma < s\, \vert Q \vert
\end{equation}
\noindent
Where $g$ and $s$ are defined in (\ref{ggs}). We also have to check the
sign of $\Delta$.

\begin{equation}
(a+1)\Delta={1\over a+1} ({a\, \Sigma^{2} \over e_{2}^{2}}+{4\, M^{2}
\over  e_{1}^{2}}) -{Q^{2}\over e_{3}^{2}} \label{esi}
\end{equation}

\noindent
Here the sign of $\Delta$ puts a constraint on the physical variables.

{\bf Type 3}

\begin{eqnarray}
\nu^{2}&=&0 \nonumber   \\
\psi&=&-\mu\,-{1\over \ln\rho+c_{2}}
\end{eqnarray}

\begin{equation}
\int {\psi(\rho)\over \rho}d\rho=-\ln(\ln\rho+c_{2})-\mu\ln\rho+c_{3}
\end{equation}

\begin{equation}
e^{\alpha_{e}\phi}=\left[{c_{2}\over \rho^{\mu} (c_{2}+
\ln\rho)}\right]^{k_{1}}
\end{equation}

\begin{equation}
C^{d}=\left({c_{2}+\ln\rho \over c_{2}}\right)^{k_{2}}[(r^{d}-
r_{1}^{d}) (r^{d}-r_{2}^{d})]^{{1\over 2}}
\rho^{{-\mu k_{1}\over 2}}
\end{equation}

\noindent
Physical parameters are found using (\ref{fiz})

\begin{eqnarray}
2M&=&(a\mu-{1\over c_{2}})\, e_{1} \nonumber  \\
\Sigma&=&(-\mu-{1\over c_{2}}) \,e_{2}   \\
Q&=&{e_{3}\over 2c_{2}} \nonumber
\end{eqnarray}

\noindent
The solution is

\begin{eqnarray}
\mu&=&-T_{2} \nonumber  \\
-{1\over c_{2}}&=&T_{1} \\
\end{eqnarray}

\noindent
which gives

\begin{equation}
{2\,Q\over e_{3}}=T_{1}
\end{equation}

\noindent
This is the equality case of the inequality (\ref{inq})

{\bf CASE 2}

\begin{equation}
\Delta=0
\end{equation}

\noindent
Then there is one root to the equation (\ref{qua}).
Denote it by $r_{1}^{d}$.

\begin{equation}
A\,C^{d}=(r^{d}-r_{1}^{d})
\end{equation}

\begin{eqnarray}
A^{2}&=&{(r^{d}-r_{1}^{d})^{2} \over C^{2d}}\mbox{\hspace{1cm}} B^{2}=
{r^{2d-2}\,C^{2} \over (r^{d}-r_{1}^{d})^{2}} \\
C^{d}&=&(r^{d}-r_{1}^{d})\,\rho^{-{\mu\,k_{1}\over 2}}\left[\cos(c_{2}+
\nu\ln\rho)\over \cos(c_{2})\right]^{k_{2}}
\end{eqnarray}

\noindent
Scalar field is given as

\begin{equation}
e^{\alpha_{e}\, \phi}=\left [{\cos{c_{2}} \over \rho^{\mu}\,\cos(c_{2}+
\nu\ln\rho)} \right ]^{k_{1}}
\end{equation}

{\bf CASE 3}
\begin{equation}
\Delta<0
\end{equation}

\noindent
This case is similar to the previous one.

\begin{equation}
e^{\alpha_{e}\, \phi}=\left [{\cos{c_{2}} \over \rho^{\mu}\,\cos(c_{2}+
\nu\ln\rho)} \right ]^{k_{1}}
\end{equation}

\begin{eqnarray}
A^{2}&=&{(r^{2d}-2b_{1}r^{d}+b_{2}) \over C^{2d}}\mbox{\hspace{1cm}} B^{2}=
{r^{2d-2}\,C^{2} \over (r^{2d}-2b_{1}r^{d}+b_{2})} \\
C^{d}&=&(r^{2d}-2b_{1}r^{d}+b_{2})^{{1\over 2}}\,\rho^{-{\mu\,k_{1}\over 2}}
\left[\cos(c_{2}+\nu\ln\rho)\over \cos(c_{2})\right]^{k_{2}}
\end{eqnarray}

\noindent
In cases $2$ and $3$, the relation of physical parameters to integration
constants are exactly the same as case 1  type 2. In addition we have the
equation
(\ref{esi}) with the sign of $\Delta$ chosen according to the case.

It is perhaps noticed that we solve field equations exactly without
assuming any special ansatz \footnote{After this work was
completed we became aware of a recent paper \cite{POL} , which
similarly solves the field equations (\ref{EJ})-(\ref{EE})
without any special ansatz, but in completely different to those used here}
. In that respect we have all the
possible integration constants in the solutions , hence
we have the following assertion:
In $D$-dimensions the metric given in (\ref{l11}) and (\ref{ME3})
is the only asymptotically flat metric corresponding
to a spherically symmetric static black hole carrying mass $M$ ,
dilaton charge $\Sigma$
and electric charge $Q$ and covered by a regular horizon.
In $D$-dimensions the Gibbons-Maeda metric is diffeomorphic to the region
$r>r_{1}$ where $r_{1}$ is the location of the outer horizon. Relaxing
the condition of asymptotical flatness there is a possibility having
a black hole (with a regular horizon and singularity located at the origin)
solution of the Einstein Maxwell Dilaton field equations \cite{MAN}

Expressing the boundary conditions of the black hole solution (\ref{ME3})
(the behaviour of the metric as $r$ approaches to the outer horizon $r_{1}$
and to infinity) as the black hole boundary conditions
in static axially symmetric four dimensions one can prove
that GHS or Gibbons-Maeda solution describes also the unique
asymptotically flat static
black hole solution in Einstein Maxwell Dilaton Gravity. In the next section
we give the proof of this statement.

\section{Uniqueness of Black Hole Soltions}

In this section we are interested in the black hole solutions
of the Einstein Maxwell Dilaton theory in the static axially
symmteric spacetimes in four dimensions. Using a different approach
the uniqueness of static charged dilaton black hole
(GHS+GM black hole) has been recently shown in ref.\cite{MAS}.
In this proof the dilaton coupling constant was taken fixed
($a=2$) which corresponds to the low energy limit of the string
theory. Here we
shall show that the static black holes of the Einstein Maxwell
Dilaton theory are unique for arbitrary values of the dilaton
coupling constant $a$.
We are not going to solve the corresponding
field equations but first formulate these field equations as
a sigma model in two dimensions and use this formulation in the proof
of the uniqueness of the solutions under the same boundary condtions.
Uniqueness of the stationary black hole solutions of the Einstein theory
is now a very well established concept \cite{MAS}-\cite{GRS}.
This proof is based on the sigma model formulation of the stationary axially
symmteric Einstein Maxwell field equations \cite{MAZ}-\cite{ERI}.
Here we shall follow the approach given by \cite{MAZ} and \cite{GRS}.

The line element of a static axially symmetric four dimensional spacetime
is given by

\begin{equation}
ds^{2}=e^{2\, \psi}\,[e^{2\gamma}\,(d\rho^{2}+dz^{2})+\rho^{2}\,d\phi^{2}\,]-
e^{-2\, \psi}\,dt^{2} ,\label{MET}
\end{equation}

\noindent
The field equations of the Einstein Maxwell Dilaton field theory with
the above metric and $A_{\mu}=(A,0,0,0)$ are

\begin{equation}
\nabla^{2}\, \psi+ {\kappa^{2} \over  2}\, e^{2\, \psi - a\, \phi}\,
(\nabla\,A)^{2}=0 \label{EQ1}
\end{equation}

\begin{equation}
\nabla^{2}\, \phi-a\, {\kappa^{2} \over  8}\, e^{2\, \psi - a\, \phi}\,
(\nabla\,A)^{2}=0 \label{EQ2}
\end{equation}

\begin{equation}
\nabla^{2}\, A + \nabla(2\psi-a\,\phi)\, \nabla\,A=0  \label{EQ3}
\end{equation}

\begin{equation}
{1 \over  \rho}\,\gamma_{,z}=2\, \psi_{,\rho}\, \psi_{,z}+
4\,\phi_{\rho}\, \phi_{,z} -\kappa^{2}\,e^{2\, \psi -a\, \phi}\,A_{\rho}\,
A_{,z}
\end{equation}

\begin{equation}
{2 \over  \rho}\,\gamma_{,\rho}=2\, \psi_{,\rho}^{2}-2\, \psi_{,z}^{2}+
4\,\phi_{,\rho}^{2}-4\, \phi_{,z}^{2} -\kappa^{2}\,e^{2\, \psi -a\, \phi}
\,(A_{,\rho}^{2}-\,A_{,z}^{2})
\end{equation}

\noindent
let $E=\psi-{a \over  2}\phi$ and $B={\kappa^{\prime} \over  \sqrt{2}}\,A$ we
then find

\begin{equation}
\nabla^{2}\, E+ \,e^{2\,E}\,(\nabla\,B)^{2}=0 \label{EQ6}
\end{equation}

\begin{equation}
\nabla^{2}\, B + 2\,\nabla\,E\, \nabla\,B=0  \label{EQ7}
\end{equation}

\noindent
where

\begin{equation}
\kappa^{\prime}=\kappa\,\sqrt {1+{a^{2} \over  8}}
\end{equation}

\noindent
We wish to write the Eqs.(\ref{EQ6} , \ref{EQ7}) as a single complex
equation by introducing a complex potential. In order to achieve this
we introduce pseudopotential $\omega$ by use of (\ref{EQ7})

\begin{equation}
\omega_{\rho}=\rho\,e^{-2E}\, B_{z} ~~ ,~~
\omega_{z}=-\rho\,e^{-2E}\, B_{\rho}
\end{equation}

\noindent
then the resulting equations can be written as the following
single complex equation (the Ernst equation) for
$\varepsilon=\rho\,e^{E}+i\omega$

\begin{equation}
Re(\varepsilon)\, \nabla^{2}\, \varepsilon=
\nabla \, \varepsilon\, \nabla\, \varepsilon
\label{E10}
\end{equation}

\noindent
Hence the above complex equation reperesents
the Eqs.(\ref{EQ6}) and (\ref{EQ7}) if we let
$\varepsilon=\rho\,e^{E}+i\,\omega$.
The above Ernst equation (\ref{E10}) defines a sigma model on $SU(2)/U(1)$
with the equation of motion

\begin{equation}
\vec{\nabla}\, \left(P^{-1}\, \vec{\nabla}\,P \right)=0 \label{SM}
\end{equation}

\noindent
where

\begin{equation}
P={1\over \rho e^{E}} \left( \begin{array}{cc} 1&-B \\ -B&\rho^{2}e^{2E}+B^{2}
\end{array} \right)
\end{equation}

In the sequel we assume enough differentiability for the components
of the matrix $P$ in $V\, \cup \, \partial V$.
Here $V$ is a region in $M$ with boundary $\partial V$. In our case
$V$ is the region $r >0$ (see section 2 type I solutions) and hence
$\partial V$ has two disconnected components.

We also assume
that $P$ is positive definite. Let $P_{1}$ and $P_{2}$ be two
different solutions of (\ref{SM}). The difference of their
equations satisfy

\begin{equation}
\vec{\nabla}\,\Bigl( P_{1}^{-1} (\vec{\nabla}\,Q )\, P_{2}\Bigr) = 0.
\label{S1}
\end{equation}

\noindent
where $Q=P_{1}\,P_{2}^{-1}$. Multiplying the both sides by $Q^{\dagger}$
(hermitian conjugation) and taking the trace we obtain

\begin{equation}
\vec{\nabla}\,\Bigl[
tr \left(Q^{\dagger}\, P_{1}^{-1}
(\vec{\nabla} Q) \, P_{2}\right)  \Bigr] =
tr \Bigl[(\vec{\nabla} Q^{\dagger})\,P_{1}^{-1}
\,(\vec{\nabla} Q)\,P_{2}\Bigr]  \label{S2}
\end{equation}

\noindent
The left hand side of the above equation can be simplified further
and we obtain

\begin{equation}
\nabla^{2}\, q =
tr \Bigl[(\vec{\nabla} Q^{\dagger} )\,P_{1}^{-1}
(\vec{\nabla} Q)\,P_{2}\Bigr]  \label{S3}
\end{equation}

n
where $q=tr(Q)$. Using the hermiticity and positive definiteness properties
of the matrices $P_{1}$ and $P_{2}$ we may let

\begin{equation}
P_{i}=A_{i}\,A_{i}^{\dagger} , \,\, (i=1,2)   \label{S4}
\end{equation}

\noindent
where $A_{1}$ and $A_{2}$ are nonsingular $n \times n$ matrices given by

\begin{equation}
A_{i}={1\over \sqrt{\rho} e^{E_{i} \over  2}} \left( \begin{array}{cc} 1&0
\\ -B_{i}&\rho e^{E_{i}} \end{array} \right)
\end{equation}

With
the aid of (\ref{S4} ) Eq.(\ref{S3}) reduces to

\begin{equation}
\nabla^{2}\, q =
tr (\vec{J}^{\dagger}\,\vec{J})
\label{S5}
\end{equation}

\noindent
where

\begin{equation}
\vec{J}= A_{1}^{-1} (\vec{\nabla} Q) \,A_{2}  \label{S6}
\end{equation}

Eq.(\ref{S5}) is a crucial step towards the proof of the uniqueness theorems.
It is clear that the right hand side is positive definite at all
point of $V$.
Before going on let us give  the scalar function $q$.

\begin{equation}
q=2+
{1 \over {\rho^{2}}\,e^{E_{1}+E_{2}}}
\left[\rho^{2}\,(e^{E_{1}}-e^{E_{2}})^2\,+\,( B_{1}\,-\, B_{2} )^{2} \right]
\label{S7}
\end{equation}

\noindent
It is clear that $q=2$ and its first derivatives vanish on the boundary
$\partial V$ of $V$.

Let $M$ be an $2$ dimensional manifold with local coordinates $(\rho,z)$.
Let $V$ be a region in $M$ with boundary $\partial V$.
Let $P$ be a hermitian positive definite $2 \times 2$ matrix with unit
determinant and let $P_{1}$ and $P_{2}$ be two such matrices satisfying
(\ref{SM}) in $V$ with the same boundary conditions on $\partial V$
then we have $P_{1}=P_{2}$ at all points in region $V$ .
The proof is as follows. Integrating (\ref{S5}) in $V$ we obtain

\begin{equation}
\int_{\partial V}\,
\vec{\nabla}\, q\,d\,\vec{\sigma} =
\int_{V}\, tr\,\bigl(\vec{J}^{\dagger}\,\vec{J}\bigr)\, dV
\label{I1}
\end{equation}

\noindent
and using the boundary condition $q=2$ on $\partial V$ we get

\begin{equation}
\int_{V}\, tr\, \bigl( \vec{J}^{\dagger}\,\vec{J}\bigr)\, dV=0
\label{I2}
\end{equation}

\noindent
Then the integrand in (\ref{I2}) vanishes at all points in $V$.
This implies
the vanishing of $\vec{J}$ which implies that $Q=Q_{0}=$
a constant matrix
in $V$. Since $Q$ is the identity matrix $I$ on $\partial V$ then $Q=I$
in $V$. Hence $P_{1}=P_{2}$ at all points in $V$.
Another way to obtain this result
is to use (\ref{S5}) directly. The vanishing of the integrand in (\ref{I2})
implies that $q$ is an harmonic function in $V$.
Since $q=2$ on the boundary $\partial V$ of V then it must be equal
to the same constant in $V$ as well. This implies that $P_{1}=P_{2}$ in $V$.

In four dimensions the Einstein Maxwell Dilaton field theory
a static black hole should carry mass $M$ , electric charge $Q$ and
dilaton charge $\Sigma$. Such a black hole solution exist which was
found by Gibbons and Maeda for an arbitrary dilaton coupling parameter $a$.
Here the above proof implies that all those solutions with the same
black hole boundary conditions (asymptotically flat and regular horizons)
as the GM solution are the same everywhere in spacetime.

\section{Exact Solutions of String Theory}

\noindent
In this section we introduce some theorems for static spherically
symmetric $D$ dimesional spacetimes which will be utilized
to obtain exact solutions of string theory. This section is a
natural extension
of the work in ref.(\cite{MET}) to an arbitrary dimensions. We shall not
give the proofs of the theorems. The necessary tools for
the proofs are given in the second section.

The covariant derivatives of $H_{ij\cdots k}$ and $k_{i}$ given in
(\ref{xy1}) and (\ref{xy2}) are expressed only
in terms of themselves and the metric tensor. Hence we have
the following theorem:

{\bf Theorem 1} {\em Covariant derivatives of the Riemann tensor $R_{ijkl}$,
the tensor $H_{ij \cdots k}$  and the vector $k_{i}$ at any order are
 expressible only in terms of $H_{ij \cdots k}\, , \,g_{ij}\, , \,k_{i}$ }

Since contraction of $k^{i}$ with $H_{ij \cdots k}$ vanishes,
the only symmetric
tensors constructable out of $H_{ij \cdots k}\, , \,g_{ij}$
and $k_{i}$ are $M_{ij}$,
the metric tensor $g_{ij}$ and $k_{i}k_{j}$. Then the following theorem
holds:

{\bf Theorem 2} {\em Any second rank symmetric tensor constructed out of
the Riemann tensor, anti-symmetric
tensor $H_{ij\cdots k}$, dilaton field $\phi=\phi(r)$ and their
covariant derivatives is a linear combination of $M_{ij}\, , \,g_{ij}$ and $
k_{i}k_{j}$.}

Let this symmetric tensor be $E^{\prime}_{ij}$. Then we have

\begin{equation}
E^{\prime}_{ij}=\sigma_{1}M_{ij}+\sigma_{2}g_{ij}+\sigma_{3}k_{i}k_{j}
\end{equation}

\noindent
where $\sigma_{1},\,\sigma_{2}$ and $\sigma_{3}$ are scalars
which are functions of the metric functions,
invariants constructed out of the curvature tensor $R_{ijkl},\,H_{ij
\cdots k} $ and the dilaton field $\phi(r)$ and their covariant derivatives.

{\bf Theorem 3} {\em Any vector constructed out of the Riemannian tensor
$R_{ijkl},\;H_{ij...k} $ the dilaton field $\phi=\phi(r)$ and their
covariant derivatives is proportional to $k_{i}$.}

\noindent
Let this vector be $E^{\prime}_{i}$. Hence
\begin{equation}
E^{\prime}_{i}=\sigma k_{i}
\end{equation}

\noindent
where $\sigma$ is a scalar like $ \sigma_{1},\;\sigma_{2},\;\sigma_{3}$.

\noindent
The covariant derivatives of $F_{ij}$, $k_{i}$ and $t_{i}$ are expressed
in terms of themselves, metric tensor. So we have a similar theorem:

{\bf Theorem 4} {\em Covariant derivatives of the Riemann tensor $R_{ijkl}$,
the antisymmetric
tensor $F_{ij}$, the vectors $k_{i}$ and $t_{i}$ at any order are
expressible only in terms of $F_{ij},\,k_{i},\,t_{i} \; and\; g_{ij}$.}

As an example to an exact solution of the low energy limit of the string
theory in D-dimensions with constant dilaton field, we propose the
LCBR (Levi-Civita Bertotti-Robinson) metric which is given by

\begin{equation}
g_{ij}={{q^{2}}\over{r^{2}}} \,(-t_{i}t_{j}
+c_{0}^{2}k_{i}k_{j} -r^{2}h_{ij}) \label{d16}
\end{equation}

\noindent
where $h_{ij}$ is the metric on $S_{D-2}$, $t_{i}=\delta_{i}^{t}$ ,
$k_{i}=\delta_{i}^{r}$ , $q$ and $c_{0}$ are constants. Then  the
antisymmteric tesor $F_{ij}$ defined in (\ref{d12}) becomes

\begin{equation}
F_{ij}={{c_{0}}\over{r^{2}}} (t_{i}k_{j}-t_{j}k_{i})
\end{equation}

\begin{equation}
\widetilde{M_{ij}}=F_{mj}F_{i}^{m}-{1\over{4}}F^{2}\,g_{ij}
\end{equation}

$e_{1}=e_{3}=0$ $\Rightarrow \,S_{ij}=e_{0} \widetilde{M_{ij}}$,
$e_{0}=q^{2}$

\begin{equation}
R_{ijkl}=q^{2}[g_{jl} \widetilde{M_{ik}}-g_{jk}\widetilde{M_{il}}+
g_{ik}\widetilde{M_{jl}}-
g_{il}\widetilde{M_{kj}}]+{{q^{2}(c_{0}^{2}-1)}\over{c_{0}^{2}}}F_{ij}F_{ml}
\end{equation}

\begin{equation}
R_{ij}=q^{2}[(D-3)+{{1}\over{c_{0}^{2}}}] \widetilde{M_{ij}}
+{{1}\over{2 q^{2}}}[(D-3)-{{1}\over{c_{0}^{2}}}]g_{ij} \label{d13}
\end{equation}

\begin{equation}
G_{ij}=q^{2}[(D-3)+{{1}\over{c_{0}^{2}}}] \widetilde{M_{ij}} +
{{1}\over{2 q^{2}}}[{{1}\over{c_{0}^{2}}}-(D-3)^{2}]g_{ij} \label{d14}
\end{equation}

\noindent
It is easily seen that

\begin{equation}
\nabla_{l}F_{ij}=0 \,\,,\,\,\nabla_{m}R_{ijkl}=0
\end{equation}

\noindent
Here the origin of the antisymmteric tensor $F_{ij}$ is geometrical. It
is the volume form of the two dimensional part $ds^2=-A^2\,dt^2+B^2\,dr^2$
of the $D$-dimesional spacetime line element (\ref{met}). We can define the
electromagnetic tensor field , $F^{el}_{ij}$ , simply as
$F^{el}_{ij}=Q\,F_{ij}$. Here $Q$ is a constant (in asymptotically flat
cases $Q$ plays the role of electric charge). Eq.(\ref{d14}) becomes

\begin{equation}
G_{ij}={q^{2}\over Q^2}\,[(D-3)+{{1}\over{c_{0}^{2}}}] \,M^{el}_{ij} +
{{1}\over{2 q^{2}}}[{{1}\over{c_{0}^{2}}}-(D-3)^{2}]g_{ij} \label{d15}
\end{equation}

\noindent
where $M^{el}_{ij}\,\,(=Q^2\,\, \widetilde{M_{ij}})$ is the energy
momentum tensor of the electromagnetic field
tensor $F^{el}_{ij}$.

In $D$ dimensions LCBR metric is the solution of the Einstein Maxwell
equations with a cosmological constant. This constant vanishes when
$(D-3)={{1}\over{c_{0}^{2}}}$. In this case spacetime is conformally flat
only when $D=4$.
For this metric, the field equations of a most general lagrangian
(for instance, the lagrangian of the low energy limit of the string
theory at all orders in string tension parameter) yields

\begin{eqnarray}
E_{ij}&=&\sigma_{2}g_{ij}+\sigma_{1}M_{ij}=G_{ij} ,\label{d20}\\
E_{i}&=&\sigma k_{i} ,\label{d21}\\
E&=&\sigma^{\prime} ,\label{d22}
\end{eqnarray}

\noindent
where $\sigma_{1}$, $\sigma_{2}$, $\sigma$ and $\sigma^{\prime}$
are constants. This means that,
Einstein's equations reduce to algebraic equations. The lagrangian
of the most general theory is a scalar containing the curvature
tensor , metric tensor , matter fields and their derivatives ,
contractions and multiple products of all orders. According to the
theorem 2 given above all second rank symmetric tensors constructed
out of these are expressible in terms of $g_{ij}$ , $M_{ij}$ , and
$k_{i}k_{j}$ . So whatever the theory is
we have three equations (under our assumptions
all the terms containing $k_{i}$ drop out) for five constants.
Two constants ($c_{0},q$) come from the definition of the
$D$-dimensional LCBR metric (\ref{d16}). One constant $Q$ comes from
the definition of the electromagnetic field tensor $F^{el}_{ij}$.
The other two constants are the constant dilaton field $\phi_{0}$
and the cosmological constant $\lambda$. In some cases $\lambda$ is set
to zero. In these cases as well, the number of equations are less
the number of unknown constants. In the general case , under our assumptions
the field equations reduce to the following algebraic equations
for $q,c_{0},Q, \phi_{0}$ , and $\lambda$

\begin{equation}
{q^2 \over Q^2}\, (D-3+ {1 \over c_{0}^2})=
\sigma_{2}(q,c_{0},Q,\phi_{0},\lambda)
\end{equation}

\begin{equation}
{1 \over 2\,q^2}\, (-(D-3)^2+ {1 \over c_{0}^2})=
\sigma_{4}(q,c_{0},Q,\phi_{0},\lambda)
\end{equation}

\begin{equation}
0\,\,=\,\,\sigma(q,c_{0},Q,\phi_{0},\lambda)
\end{equation}

\begin{equation}
0\,\,=\,\,\sigma^{'}(q,c_{0},Q,\phi_{0},\lambda)
\end{equation}

\noindent
where $\sigma_{1}$, $\sigma_{2}$, $\sigma$ and $\sigma^{\prime}$
are defined in (\ref{d20})-(\ref{d22}). Choosing $\phi$ ($=\phi_{0}$) and
the metric function
$C$ ($=c_{0}$ as constants the function $\sigma$ given
above
equations vanishes identically. Hence we are left with three algebraic
equations for $q,c_{0},Q, \phi_{0}$ , and $\lambda$.
These algebraic equations may or may not have solutions. For intance
when we consider the Lovelock theory \cite{LOV} without
a cosmological contant these equations have
solutions when $D=4$ and $D>6$. With a cosmological constant these
algebraic equations have solutions for $D \ge 4$.
In the case of the string theory LCBR metric is a solution of the
low energy limit of the string theory.
LCBR spacetime in four dimensions is also known to be an exact solution of
the string theory \cite{LOW} , \cite{MET}.
Here we have the generalization of this result for an arbitrary value of
$D>2$.
If there exists a
solution of the above mentioned algebraic equations then

{\bf Theorem 5} {\em LCBR metric is an exact solution of the
string theory at all orders }

As in the case of the LCBR spacetime \cite{LOW} the symmteric spaces ,
$\nabla_{i}\,R_{jklm}=0$ and reccurrent spaces ,
$\nabla_{i}\,R_{jklm}=l_{i}\, R_{jklm}$ where $l_{i}$
is a (gradient) vector , play effective role in constructing exact
solutions of the string theory. These spaces are in general product
spaces \cite{SC}. In the case of static spherically symmetric spacetimes
the only possibility to have such spaces is to choose $C=c_{0}$. Then
$D$ dimensional spacetime becomes $M_{2} \times S^{D-2}$ , where $M_{2}$
is a two dimesional geometry with metric
$ds^{2}=-A^{2}\,dt^{2}+B^{2}\,dr^{2}$.

\noindent
If we take $C=c_{0}$ in the metric (\ref{met}), then

\begin{eqnarray}
\eta_{0}&=&{c_{0}^{2(D-2)}\over (D-3)!}\, {1\over AB} \left(A^{\prime}\over
B\right)^{\prime} \nonumber \\
\eta_{1}&=&0  \\
\eta_{2}&=&\xi-(D-3)\eta_{0} \nonumber \\
\eta_{3}&=&0 \nonumber
\end{eqnarray}

\noindent
where $\xi={c_{0}^{2(D-3)}\over (D-4)!}$.

\begin{eqnarray}
R_{ij}&=& {(D-3)!\over 2 c_{0}^{2 (D-2)}}\,\,
[-\eta_{0}+\xi]\,g_{ij}+[\eta_{0}+\xi]\,M_{ij} \nonumber \\
R&=&{(D-3)!\over c_{0}^{2 (D-2)}} \,\,[-2\eta_{0}+(D-2)\xi] \\
G_{ij}&=&- {(D-3)!\over 2 c_{0}^{2 (D-2)}}\,\,
[-\eta_{0}+(D-3)\xi]\,g_{ij}+[\eta_{0}+\xi]\,M_{ij}  \nonumber
\end{eqnarray}

\begin{eqnarray}
\nabla_{l}H_{ij\ldots m}&=&0 \nonumber \\
\nabla_{i}k_{j}&=&{A^{\prime}\over 2AB^{2}}\,g_{ij}+{c_{0}^{2(D-2)}\over
(D-3)!}{A^{\prime}\over AB^{2}}M_{ij}-{(AB)^{\prime}\over AB}k_{i}k_{j}
\end{eqnarray}

\noindent
If we let $\eta_{0}$ to be constant, then we have to solve the equation
${1\over AB} \left(A^{\prime}\over B\right)^{\prime}=\alpha$.
The solution is

\begin{equation}
B^{2}={A^{\prime 2}\over \alpha A^{2}+\beta} \label{bbb}
\end{equation}

\noindent
where $\alpha$ and $\beta$ are constants. Hence the metric
 (\ref{met}) can be written as

\begin{equation}
ds^{2}=-A^{2}dt^{2}+{dA^{2}\over \alpha A^{2}+\beta}+
c_{0}^2\,d\Omega_{D-2}^{2}
\end{equation}

\noindent
The case $\beta=0$ is identical with the LCBR metric.
As in the case of the LCBR metric the metric
constants $\alpha$ , $c_{0}$ and the magnetic charge satisfy three coupled
(due to the theorems given before) algebraic equations. If these
algebraic equations have solutions then the metric given
above with nonvanishing $\alpha$ and $\beta$ is a candidate as an exact
solution of the string theory.

Above we have given the Riemann
tensor in terms of the volume form of $S_{D-2}$ (D-2)-brane field.
We may also write the Riemann tensor in the form

\begin{equation}
R_{ijml}=g_{jl}S_{im}-g_{jm}S_{il}+g_{im}S_{jl}-g_{il}S_{mj}
+e_{2}F_{ij}F_{ml}
\end{equation}

\noindent
In this case

\begin{eqnarray}
e_{0}&=&c_{0}^{2} \nonumber \\
e_{1}&=&0  \\
e_{2}&=&-c_{0}^{2}+c_{0}^{2} \zeta  \nonumber  \\
e_{3}&=&0  \nonumber
\end{eqnarray}

\noindent
where $\zeta={c_{0}^{2}\over AB}\left({A^{\prime}\over B}\right)^{\prime}$

\begin{eqnarray}
R_{ij}&=&{D-3-\zeta\over 2c_{0}^{2}}\,g_{ij}+(D-3+
\zeta)c_{0}^{2}\widetilde{M_{ij}} \nonumber  \\
R&=&{D^{2}-5D+6-2\zeta\over c_{0}^{2}} \\
G_{ij}&=&{\zeta-(D-3)^{2}\over 2c_{0}^{2}}\,g_{ij}+(D-3+
\zeta)c_{0}^{2}\widetilde{M_{ij}} \nonumber \\
\end{eqnarray}

\begin{eqnarray}
\nabla_{l}F_{ij}&=&0 \nonumber \\
\nabla_{i}k_{j}&=&{A^{\prime}\over 2AB^{2}}\,g_{ij}+
c_{0}^{4}{A^{\prime}\over AB^{2}}\widetilde{M_{ij}}-
{(AB)^{\prime}\over AB}k_{i}k_{j}
\end{eqnarray}

\noindent
The solution given in (\ref{bbb}) is not valid
when $A$ is set to a constant. In such a case  $B$ can be choosen as unity.
Therefore letting $A=B=1$ and $C=c_{0}$. We find that
$e_{0}=-e_{2}=c_{0}^{2}$ and

\begin{equation}
\nabla_{l}F_{ij}=0 \,\,,\,\,\nabla_{m}R_{ijkl}=0
\end{equation}

\begin{equation}
\nabla_{i}k_{j}=0 \,\,,\,\,\nabla_{i}t_{j}=0
\end{equation}

\begin{eqnarray}
R_{ij}=(D-3)c_{0}^{2}\,\left[{\widetilde M}_{ij}+
{1 \over  2c_{0}^{2}}g_{ij} \right]\\
R={(D-3)(D-2) \over  c_{0}^{2}}\\
G_{ij}=(D-3)c_{0}^{2}\,\left[{\widetilde M}_{ij}+
{1 \over  2c_{0}^{2}}(3-D)g_{ij} \right]\\
\end{eqnarray}

\noindent
Since the metric contains only one arbitrary constant $c_{0}$
it may not be possible to eliminate the coefficient of $g_{ij}$
(the cosmological constant) in the field equations. Hence the
metric given by

\begin{equation}
ds^{2}=-dt^{2}+dr^{2}+c_{0}^{2}\,d\, \Omega_{D-2}^{2}
\end{equation}

\noindent
is a canditate as an exact solution of
the string equations with a cosmological constant.

\section{Conclusion}

We have found all possible asymptotically flat
solutions of Einstein Maxwell Dilaton field
theory in $D$ dimensional static spherically symmetric spacetimes
 with arbitrary
dilaton coupling constant. We proved that the asymptotically flat
black hole solutions
carrying mass $M$ , electric charge $Q$ and dilaton charge $\Sigma$
in four dimensions are unique not only in static spherically symmteric
spacetimes but also in static axially symmetric spacetimes. This proof
may be
extended to arbitrary dimensions and also to spacetimes which are not
asymptotically flat \cite{MAN}. We searched for all possible
static spherically spacetimes which may solve the string equations exactly
(to all orders in string tension parameter).

\vspace{0.5cm}
\noindent
This work is partially supported by the Scientific and Technical Research
Council of Turkey (TUBITAK) and Turkish Academy of Sciences (TUBA).
We would like to thank the referees for their several suggestions ,
in particular for informing us the recent works \cite{POL} , \cite{MAN}.

\end{document}